\documentclass[sigconf,nonacm]{acmart}

\usepackage{pgfplots}
\pgfplotsset{compat=1.18}

\AtBeginDocument{%
  \providecommand\BibTeX{{%
    \normalfont B\kern-0.5em{\scshape i\kern-0.25em b}\kern-0.8em\TeX}}}

\usepackage{booktabs}
\usepackage{graphicx}      
\usepackage{amssymb}       
\newcommand{\cmark}{\checkmark}
\newcommand{\pmark}{$\circ$}

\begin{document}

\title{$\tau$-Rec: A Verifiable Benchmark for Agentic Recommender Systems}

\author{Bharath Sivaram Narasimhan}
\affiliation{%
  \institution{Independent Researcher}
  \city{Mountain View}
  \state{CA}
  \country{USA}
}
\email{nbharaths@gmail.com}

\author{Karthik R Narasimhan}
\affiliation{%
  \institution{Princeton University}
  \city{Princeton}
  \state{NJ}
  \country{USA}
}
\email{karthikn@princeton.edu}

\renewcommand{\shortauthors}{Narasimhan and Narasimhan}

\newcommand{\passhat}[1]{\textrm{pass\textasciicircum#1}}
\newcommand{\taubench}[0]{$\tau$-bench}
\newcommand{\taurec}[0]{$\tau$-Rec}

\begin{abstract}
As recommender systems transition toward agentic, multi-turn conversational interfaces, evaluation paradigms have struggled to keep pace. Current benchmarks often rely on ``LLM-as-a-judge'' evaluations, which introduce subjectivity, high costs and inconsistency. We present \textbf{$\boldsymbol{\tau}$-Rec}, a benchmark for agentic recommender systems that replaces subjective evaluation with \textbf{verifiable rewards} and a reveal-tagged elicitation (RTE) mechanism that controls how task constraints surface during dialogue. By testing agents against structured catalog predicates and employing a \textbf{\passhat{k}} reliability metric, $\tau$-Rec provides a systematic test for consistent reasoning. Our evaluation of nine configurations across five model families — GPT-5.4, Claude Sonnet 4.6, Gemini 2.5 Flash, DeepSeek V4 Flash, Qwen3-32B and GPT-5 mini — reveals a steep reliability cliff, where even the best model achieves only $\sim$57\% at \passhat{1} and $\sim$35\% at \passhat{4}, highlighting a critical gap in current conversational agent deployment. All code and data are publicly available at \textbf{\url{https://github.com/nbharaths/tau-rec}}.
\end{abstract}

\maketitle

\section{Introduction}
Modern recommender systems are evolving from static, single-turn ranking pipelines into agentic systems where Large Language Models (LLMs) drive multi-turn dialogues, invoke external tools to gather context, and elicit user preferences progressively through conversation. Recent systems such as InteRecAgent \cite{huang2025interecagent}, RecMind \cite{wang2024recmind}, MACRec \cite{wang2024macrec}, and AgentRecBench \cite{shang2025agentrecbench} illustrate this trajectory in which agents now plan, query catalogs, reason over multi-criteria constraints, and adapt within a session. 

However, evaluation of agentic recommender systems has not kept pace. Existing benchmarks fall into two camps, both inadequate for agentic settings. First, the static-dialogue camp, including works such as INSPIRED \cite{hayati2020inspired}, DuRecDial \cite{liu2020durecdial}, OpenDialKG \cite{moon2019opendialkg}, and CRSLab \cite{zhou2021crslab}, measures recommendation quality with surface-level metrics like BLEU and Recall@k against fixed annotated dialogues, making them vulnerable to memorization~\cite{dipalma2025memorize,zhang2026leakage}. Second, the LLM-as-judge camp, including MT-Bench \cite{zheng2023mtbench}, UserSimCRS v2 \cite{bernard2026usersimcrs} and CRS Arena \cite{bernard2025crsarena}, relies on subjective scoring that is expensive, non-deterministic, and inconsistent across runs. Bernard \& Balog \cite{bernard2025metrics} further documented that existing metrics for conversational recommender systems (CRS) correlate only weakly with real user satisfaction.

We propose \taurec{}, a verifiable benchmark for agentic conversational recommendation. \taurec{} frames recommendation as a constrained optimization problem in a multi-turn dialogue and treats the interaction as a Partially Observable Markov Decision Process (POMDP) over a tool-agent-user (TAU) loop. The recommender agent has access to catalog and metadata tools, while a user simulator holds private preferences that the agent must elicit through conversation. The agent is tested simultaneously on two competencies: (i) preference elicitation — asking the right questions to surface unstated user requirements, and (ii) constrained reasoning — invoking the right tools and combining their results to satisfy the elicited constraints. Building on \taubench{} \cite{yao2024taubench}, we adopt the \passhat{k} metric, which measures the probability of solving a task correctly across all $k$ independent trials.\footnote{Although our default scoring requires every dimension to be satisfied for a success, we also analyze partial-credit failure modes to attribute exactly which constraint or policy dimensions an agent fails on.} \passhat{k} surfaces a reliability dimension that capability-only metrics like Recall@N and Hit Rate@N cannot detect, and brings reliability as a new evaluation paradigm to agentic recommendation. We populate the data catalog with movies from TMDB that are post training cutoff for major LLMs.
While our presentation focuses on movies, the framework is domain-agnostic and extends naturally to other domains like music, books, podcasts, or e-commerce.

\taurec{} also introduces a policy-compliance objective absent from existing CRS benchmarks. Inspired by \taubench{}'s domain-specific policy documents \cite{yao2024taubench} and by MATCHA's \cite{hui2025matcha} safety-policy work in game recommendation, our policies test not just what the agent recommends but how it recommends it, e.g., whether age-restricted content is appropriately gated, whether sponsored items are disclosed, whether already-consumed items are filtered, and whether the agent abstains on impossible requests rather than fabricating a recommendation. This dimension makes responsible-AI behavior a first-class evaluation target.

We evaluate six contemporary models —  GPT-5.4, Claude Sonnet 4.6, Gemini 2.5 Flash, GPT-5 mini, DeepSeek V4 Flash and Qwen3-32B — as agents, with GPT-5 mini as the user simulator. The results expose a striking reliability cliff: even the strongest agent achieves only $\sim$57\% at \passhat{1} and drops to $\sim$35\% at \passhat{4}, and the capability-latency Pareto frontier (Figure~\ref{fig:latency}) shows GPT-5.4 (no thinking) anchors the speed end while DeepSeek V4 Flash variants top the capability end.

\section{Related Work}
\textbf{CRS benchmarks.} Early CRS evaluation centered on static dialogue corpora — INSPIRED \cite{hayati2020inspired}, DuRecDial \cite{liu2020durecdial}, and OpenDialKG \cite{moon2019opendialkg} — consolidated under CRSLab \cite{zhou2021crslab}. These resources rely on BLEU and Recall@k against fixed reference dialogues, measuring surface similarity rather than task completion, and are exposed to contamination \cite{dipalma2025memorize,zhang2026leakage}.
The closest agentic precedent, AgentRecBench \cite{shang2025agentrecbench}, evaluates LLM-agent recommendation over Amazon, Goodreads, and Yelp, but its Hit Rate@N over 1-positive/19-negative pools is a single-turn ranking metric, not multi-turn dialogue with a simulated user — and it has no \passhat{k}, no policies, and no fresh-catalog mechanism. CRS Arena \cite{bernard2025crsarena} takes the complementary route of crowdworker Elo battles and finds even the best CRS satisfies users only $\sim$52\% of the time. LiveCodeBench \cite{jain2024livecodebench} and SciNUP \cite{scinup} establish post-cutoff content as a contamination defense in code and scholarly domains but no recommendation analog exists.
 
\textbf{Agentic and tool-use evaluation.} \taubench \cite{yao2024taubench,tau2bench} is the closest methodological predecessor to \taurec, evaluating tool-agent-user interaction with deterministic database-state verification, domain-specific policy documents, LLM-driven user simulators, and \passhat{k}. Recommendation introduces fundamentally different challenges: preference elicitation under uncertainty, catalog exploration over a large item space, multi-criteria optimization, and incremental intent revelation, requiring different tool APIs, different policy types, and a different success criterion. ToolBench \cite{qin2023toolbench}, AppWorld \cite{trivedi2024appworld}, and IFEval \cite{zhou2023ifeval} also explore verifiable tool-use or instruction-following criteria but outside the recommendation setting. On the systems side, InteRecAgent \cite{huang2025interecagent}, RecMind \cite{wang2024recmind}, and MACRec \cite{wang2024macrec} build tool-augmented CRS architectures but evaluate themselves on Recall@K and NDCG@K over existing datasets without standardized tasks, policies, or \passhat{k}. MATCHA \cite{hui2025matcha} is the closest existing work to \taurec's policy dimension — it incorporates explicit safety/risk-control policies and reports a 97.9\% adversarial defense rate — but it is a deployed system on a proprietary dataset, not a reusable benchmark. RecBot \cite{alibaba2025recbot} reports a "pass rate" over a 3-month A/B test, conceptually related to but less rigorous than \passhat{k}.
 
\textbf{User simulation.} iEvaLM~\cite{wang2023ievalm} established LLM-driven user simulation for CRS but uses memorization-prone datasets, scores with Recall@k, and requires no tool use or policies. UserSimCRS v2~\cite{bernard2026usersimcrs} supports agenda and LLM-based simulators with structured information needs, but assumes text-generating agents and treats policies as out of scope. ConvApparel~\cite{meshi2026convapparel} and RecUserSim~\cite{recusersim} advance simulator validation without coupling to verifiable downstream tasks. None combine LLM role-play with \textbf{reveal-tagged constraints} — the mechanism by which \taurec{} forces agents to elicit information through dialogue rather than pattern-match against a disclosed preference list. Table~\ref{tab:crs-comparison} summarizes how \taurec{} combines features that prior resources provide only individually.


\begin{table}[t]
\centering
\caption{Comparison of CRS evaluation resources. \cmark{} = fully supported, \pmark{} = partial. VR=verifiable rewards, MT=multi-turn dialogue, TU=tool use, PC=policy checks, FC=fresh catalog, HI=hidden-intent simulation.}
\label{tab:crs-comparison}
\small
\setlength{\tabcolsep}{5pt}
\renewcommand{\arraystretch}{1.1}
\begin{tabular}{l ccccccc}
\toprule
\textbf{Resource}
  & \textbf{VR}
  & \textbf{MT}
  & \textbf{TU}
  & \textbf{PC}
  & \textbf{FC}
  & \textbf{$\passhat{k}$}
  & \textbf{HI} \\
\midrule
CRSLab~\cite{zhou2021crslab}                   &        & \pmark &        &        &        &        &        \\
iEvaLM~\cite{wang2023ievalm}                   &        & \cmark &        &        &        &        & \pmark \\
UserSimCRS v2~\cite{bernard2026usersimcrs}     & \pmark & \cmark &        &        &        &        & \pmark \\
AgentRecBench~\cite{shang2025agentrecbench}    & \pmark &        & \cmark &        &        &        &        \\
InteRecAgent~\cite{huang2025interecagent}      &        & \cmark & \cmark &        &        &        & \pmark \\
MATCHA~\cite{hui2025matcha}                    &        & \cmark & \cmark & \pmark &        &        &        \\
\midrule
\textbf{$\tau$-Rec (ours)}
  & \textbf{\cmark} & \textbf{\cmark} & \textbf{\cmark}
  & \textbf{\cmark} & \textbf{\cmark} & \textbf{\cmark} & \textbf{\cmark} \\
\bottomrule
\end{tabular}
\end{table}

\section{\texorpdfstring{$\tau$-Rec}{tau-Rec} Benchmark Design}

\taurec{} models the recommendation interaction as a Partially Observable Markov Decision Process (POMDP) over a \emph{tool--agent--user} (TAU) triad. At each turn, the recommender agent observes (i) the dialogue history with a simulated user, (ii) the results of any tool calls it has issued so far, and (iii) the catalog metadata exposed through tool responses. The user's full constraint set---including hidden constraints---is unobservable and must be inferred through dialogue. The agent's action space is the union of (a) natural-language utterances directed at the user, (b) catalog tool invocations (search, filter, metadata lookup, availability check), and (c) a terminal \texttt{recommend} tool that submits a single candidate. The episode terminates when the agent issues a recommendation or a turn budget is exhausted.

The environment rests on four technical pillars, each motivated by a specific failure mode of prior CRS evaluation.

\textbf{Verifiable rewards.} Success is checked against typed catalog predicates such as \texttt{runtime <= 120} or \texttt{content\_rating} $\in \{\text{PG-13},\text{G}\}$, all non-subjective judgments. The primary reward is the product \texttt{constraint\_score} $\times$ \texttt{policy\_score}, both in $[0,1]$. Under the strict scoring used in our experiments, a recommendation succeeds only when every required constraint dimension is satisfied and every active policy is respected. We additionally report partial-credit decompositions to attribute failures to specific dimensions. Because scoring evaluates typed predicates directly against the catalog, it requires zero LLM calls and is fully deterministic and reproducible.

\textbf{Reveal-tagged elicitation (RTE).} Each constraint in a task is tagged as \texttt{volunteer} (the user states it proactively in the opening turn), \texttt{on\_ask} (the user states it only when explicitly asked about that attribute), or \texttt{hidden} (the user never states it explicitly but rejects recommendations that violate it). This tagging forces the agent to elicit information through dialogue rather than receive a complete preference dump in the first turn, closing the loophole that has made many existing CRS benchmarks easy to game by single-shot pattern matching.

\textbf{\passhat{k} reliability metric.} Following \taubench{} \cite{yao2024taubench}, we report \passhat{k}, the probability that the agent solves a task on all $k$ independent trials, rather than mean success across trials.
\begin{displaymath}
\passhat{k} = \frac{1}{|T|} \sum_{t \in T} \frac{\binom{c_t}{k}}{\binom{n_t}{k}}
\end{displaymath}
\passhat{1} measures capability; \passhat{k} for $k>1$ measures consistency. As Bernard \& Balog \cite{bernard2025metrics} argued, capability-only metrics conceal reliability cliffs that surface when an agent must succeed repeatedly. \passhat{k} is, to our knowledge, novel to recommendation evaluation.

\textbf{Policy enforcement as a first-class objective.} Each task carries a list of active policy flags drawn from a domain-specific policy document. Programmatic checks audit the trajectory for compliance with seven policies: \texttt{recommend\_tool} (the agent must issue its recommendation through the dedicated tool, not in free text), \texttt{watch\_history} (do not recommend already-consumed items), \texttt{availability} (only recommend titles available on the user's stated streaming services), \texttt{age\_restricted} (gate adult content for younger personas), \texttt{sponsored} (disclose sponsored placement), \texttt{transparency} (on impossible tasks, explicitly abstain rather than force a non-existent recommendation), and \texttt{single\_recommendation} (return a single concrete recommendation, not a list). Policy compliance is what differentiates \taurec{} from AgentRecBench \cite{shang2025agentrecbench} since we test not only what the agent recommends but how it conducts the recommendation interaction.


\subsection{Movie Catalog Construction}

The benchmark catalog is sourced from The Movie Database (TMDB) via its public REST API. Our construction pipeline has three stages:

\textbf{Stage 1: Discovery.} We query TMDB's \texttt{/discover/movie} endpoint with date filters drawing from 2025--2026 releases to ensure post-training cutoffs for all major LLMs, sorting by popularity and vote count to surface titles with sufficient metadata coverage.

\textbf{Stage 2: Enrichment.} For each candidate, we issue follow-up requests to retrieve full metadata: genre tags, runtime in minutes, MPAA-style content rating (G / PG / PG-13 / R / NR), cast and director, vote average and vote count (used as a quality proxy), release date, and per-region streaming-provider information from \texttt{/movie/\{id\}/watch/providers} (e.g., Netflix, Prime Video, Hulu, Disney+, Apple TV+).

\textbf{Stage 3: Normalization and validation.} Records are normalized to a typed schema, lightly deduplicated, and filtered to ensure that all task-relevant attributes (\texttt{runtime}, \texttt{genres}, \texttt{content\_rating}, \texttt{streaming\_services}, \texttt{rating}) are populated. Items missing any required field are dropped to guarantee that constraint predicates  can be evaluated cleanly.

The current catalog contains \textbf{153 movies}, and can easily be extended. Each catalog entry exposes the typed fields above, which together form the predicate basis for constraint verification. The pipeline is parameterized and idempotent, allowing a regular refresh to produce a new catalog snapshot pinned to a date, and all task constraints can be automatically re-validated against the new catalog to confirm task solvability remains intact.

\subsection{Task Construction}

We authored tasks using a structured protocol. Each task specifies:

\begin{enumerate}
\item A natural-language \textbf{persona} for the simulated user (e.g., a tired parent looking for a short comedy after their kids are asleep).
\item A set of typed \textbf{constraints} over the catalog schema (genres, runtime bounds, content-rating allow-list, minimum vote average, required streaming services), each with an RTE tag \texttt{volunteer} / \texttt{on\_ask} / \texttt{hidden}.
\item Optional \textbf{soft preferences} used by the simulator to style its responses (e.g., ``I prefer recent films, but it's not a hard rule''), but not scored against the agent's recommendation.
\item A list of active \textbf{policy flags} that the agent's behavior must respect over the course of the trajectory.
\end{enumerate}
We stratify the tasks along two axes. \emph{Complexity} is set by constraint count, either \textbf{simple} (1--2 constraints), \textbf{medium} (3--4), or \textbf{complex} (5+). \emph{Reveal difficulty} is set at the task level: \textbf{volunteer} tasks expose all constraints proactively, \textbf{mixed} tasks include at least one \texttt{on\_ask} constraint, and \textbf{hidden} tasks include at least one constraint the simulator never states explicitly. The current release comprises 60 tasks across a $3 \times 3$ complexity (20 simple / 24 medium / 16 complex) $\times$ reveal-difficulty (13 volunteer / 32 mixed / 15 hidden) grid, including 5 \textbf{no-valid-recommendation} tasks (where no catalog item satisfies the full constraint set) to test whether agents correctly \emph{abstain} rather than fabricate a recommendation.



\begin{table*}[!t]
\caption{Per-model reliability (\passhat{k} = probability of succeeding on \emph{all} $k$ trials), mean constraint/policy sub-scores, top policy violation rates (\texttt{avail.}=availability, \texttt{w.hist.}=watch history, \texttt{transp.}=transparency, \texttt{No-rec}=no recommendation issued), and efficiency (Turns = mean turns to recommendation, Tools = median tool calls per task). 95\% bootstrap CIs on $\passhat{4}$: $\pm$0.10--0.13.}
\label{tab:aggregate}
\small
\setlength{\tabcolsep}{4pt}
\begin{tabular}{l ccc cc cccc cc}
\toprule
 & \multicolumn{3}{c}{\textbf{Reliability}} & \multicolumn{2}{c}{\textbf{Scores}} & \multicolumn{4}{c}{\textbf{Top policy violations}} & \multicolumn{2}{c}{\textbf{Efficiency}} \\
\cmidrule(lr){2-4}\cmidrule(lr){5-6}\cmidrule(lr){7-10}\cmidrule(lr){11-12}
\textbf{Model} & \textbf{$\passhat{1}$} & \textbf{$\passhat{2}$} & \textbf{$\passhat{4}$} & \textbf{Constr.} & \textbf{Policy} & \textbf{\texttt{avail.}} & \textbf{\texttt{w.hist.}} & \textbf{\texttt{transp.}} & \textbf{\texttt{No-rec}} & \textbf{Turns} &
\textbf{Tools}\\
\midrule
GPT-5.4 (no thinking)             & 0.471 & 0.358 & 0.283 & 0.496 & 0.908 & 0.075 & 0.004 & 0.013 & 0.308 & 0.73 & 10 \\
GPT-5.4 (med thinking)            & 0.551 & 0.450 & 0.350 & 0.565 & 0.958 & 0.008 & 0.004 & 0.029 & 0.218 & 0.81 & 25 \\
Sonnet 4.6                        & 0.537 & 0.433 & 0.350 & 0.579 & 0.896 & 0.092 & 0.000 & 0.013 & \textbf{0.196} & 1.18 & 16 \\
Gemini 2.5 Flash                  & 0.275 & 0.189 & 0.138 & 0.311 & 0.920 & 0.059 & 0.000 & 0.008 & 0.672 & 0.92 & 8 \\
DeepSeek V4 Flash (no thinking)   & 0.546 & 0.433 & 0.333 & 0.583 & 0.921 & 0.067 & 0.000 & 0.013 & 0.221 & 1.47 & 28 \\
DeepSeek V4 Flash (high thinking) & 0.560 & \textbf{0.461} & \textbf{0.383} & 0.586 & 0.937 & \textbf{0.000} & 0.000 & \textbf{0.000} & 0.230 & 1.70 & 29 \\
DeepSeek V4 Flash (max thinking)  & \textbf{0.571} & \textbf{0.461} & 0.350 & \textbf{0.596} & 0.942 & \textbf{0.000} & 0.000 & \textbf{0.000} & 0.229 & 1.76 & 30 \\
Qwen3-32B                         & 0.271 & 0.181 & 0.117 & 0.361 & 0.756 & 0.210 & 0.013 & 0.008 & 0.433 & 0.33 & 5 \\
GPT-5 mini                        & 0.417 & 0.333 & 0.250 & 0.429 & \textbf{0.963} & 0.021 & 0.000 & 0.017 & 0.446 & 0.96 & 19 \\
\bottomrule
\end{tabular}
\end{table*}

\begin{table*}[!t]
\caption{$\passhat{1}$ stratified by reveal difficulty (task counts in parentheses).}
\label{tab:reveal}
\small
\setlength{\tabcolsep}{4pt}
\begin{tabular}{lccccccccc}
\toprule
\textbf{Reveal} & \textbf{GPT-5.4} & \textbf{GPT-5.4+T} & \textbf{Sonnet 4.6} & \textbf{Gem-2.5F} & \textbf{DS-V4F} & \textbf{DS-V4F+T(hi)} & \textbf{DS-V4F+T(max)} & \textbf{Qwen3-32B} & \textbf{GPT-5 mini} \\
\midrule
Volunteer (13) & 0.750 & 0.891 & \textbf{0.923} & 0.500 & 0.846 & 0.827 & 0.865 & 0.481 & 0.712 \\
Mixed     (32) & 0.477 & 0.516 & 0.523 & 0.289 & 0.586 & \textbf{0.609} & 0.586 & 0.281 & 0.414 \\
Hidden    (15) & 0.217 & \textbf{0.333} & 0.233 & 0.050 & 0.200 & 0.222 & 0.283 & 0.067 & 0.167 \\
\bottomrule
\end{tabular}
\end{table*}

\section{Benchmark Validation}
We evaluated nine configurations (six base models, with GPT-5.4 run in two thinking modes and DeepSeek V4 Flash run in three thinking modes) on all 60 tasks across 4 trials each, with GPT-5 mini as the simulator\footnote{We ran a subset of trials with various simulators and found GPT-5 mini to provide good performance at reasonable cost.} (Table~\ref{tab:aggregate}). We use \passhat{k} to measure reliability as the probability an agent succeeds on \textbf{all} $k$ trials of a task. All runs use LiteLLM with \texttt{temp=0} for agent models (where supported) and \texttt{temp=1.0} for the simulator. 

\textbf{Main results.}
GPT-5.4, Sonnet 4.6, and DeepSeek V4 Flash variants form a top tier (\passhat{1}: 0.537–0.571) with overlapping confidence intervals. All models have lower \passhat{2} and \passhat{4} scores demonstrating lack of consistency in their agentic performance. 

\textbf{Policy Compliance.}
Policy compliance varies sharply across the model lineup
(Table~\ref{tab:aggregate}, \emph{Policy} column). The strongest tier
(GPT-5.4 with thinking, GPT-5 mini, DeepSeek V4 Flash) has  >0.92 compliance, while
Qwen3-32B falls to 0.756 with a 0.21
\texttt{availability} violation rate, where the model recommends titles
without verifying streaming-service eligibility. 

\textbf{Efficiency.}
Tool use varies considerably across models,  from
DeepSeek V4 Flash's 28 median calls per trial to Qwen3-32B's 5 and Gemini 2.5 Flash's 8. The high-volume models (DS V4 Flash, GPT-5.4 with thinking, Sonnet 4.6) take more turns to recommendation but achieve higher constraint scores. 

\textbf{Failure Analysis.}
From table~\ref{tab:aggregate}, we identify three failure modes. \emph{Abstainers}
(Qwen3-32B, GPT-5 mini, Gemini 2.5 Flash) fail by not committing, having
\texttt{No-rec} rates of 0.43, 0.45, 0.67 with single-digit median tool calls;
they exit without issuing \texttt{recommend()}. \emph{Wrong-committers}
commit confidently to constraint-violating recommendations: Qwen3-32B's
0.21 availability-violation rate coupled with high commitment is the
clearest case. On the other hand, balanced models (DeepSeek V4 Flash, GPT-5.4 with
thinking, Sonnet 4.6) invest 15--28 median tool calls \emph{and} commit,
yielding the highest constraint scores (0.57--0.58). \passhat{k}
amplifies these distinctions and under repeated trials, neither hedging nor bluffing converges, and the cliff steepens.

\textbf{Difficulty gradient.}
The reveal-tag mechanism produces a steep difficulty gradient that
holds across all models (Table~\ref{tab:reveal}). For DeepSeek V4 Flash, \passhat{1} drops from
0.846 on volunteer tasks (constraints stated up front) to 0.586 on
mixed tasks (one or more \texttt{on\_ask} constraints) to 0.200 on
hidden tasks (one or more \texttt{hidden} constraints) --- a 4$\times$
gap purely from how information is revealed. The same gradient
appears for GPT-5 mini (0.712 / 0.414 / 0.167) and Qwen3-32B (0.481 /
0.281 / 0.067). Hidden constraints expose the elicitation-vs-retrieval
distinction: agents that excel at constraint-satisfaction over a known
preference set struggle when the preference must first be inferred
from rejection signals.

\begin{figure}[t]
\centering
\includegraphics[width=0.9\columnwidth]{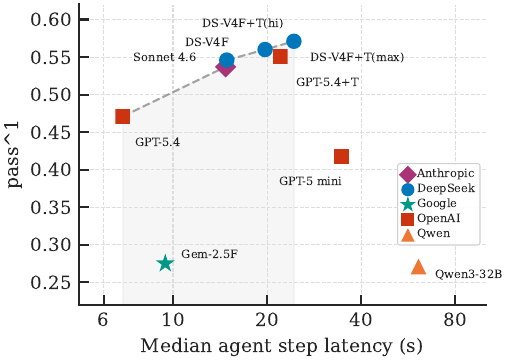}
\caption{Capability vs latency Pareto frontier. Latency = median agent step latency. Suffix +T = thinking enabled; (hi)/(max) = DeepSeek thinking budgets. DS-V4F = DeepSeek V4 Flash; Gem-2.5F = Gemini 2.5 Flash.}
\label{fig:latency}
\vspace{-12pt}
\end{figure}

\textbf{Capability-latency frontier.} 
We also analyze the capability vs latency trade-offs for different models, since per-turn latency is an important consideration for any agentic RS (Figure~\ref{fig:latency}). GPT-5.4 (no thinking)
anchors the fast end of the frontier ($\passhat{1}=0.47$ at $\sim$7s/step), while
Sonnet 4.6 ($\sim$15s) and the DeepSeek V4 Flash variants ($\sim$15--24s)
trace the rest of the frontier, topping out at $\passhat{1}=0.57$ for
DS V4 Flash (max thinking). Off-frontier models (GPT-5 mini, Qwen3-32B,
Gemini 2.5 Flash) are dominated by frontier alternatives at similar or
shorter step latency. Thinking modes follow shallow, sub-linear trajectories, e.g.
DS V4 Flash gains only +0.025 $\passhat{1}$ from non-thinking to max-effort
thinking, suggesting current models are capability constrained for this benchmark.


\section{Limitations and Future Work}
\textbf{Catalog scale.} The catalog is intentionally small (153 titles) to isolate reasoning ability from retrieval scale — agents access it via API tools, so size is opaque to the model — but limits failure-mode diversity; scaling while preserving post-cutoff freshness is future work.\\
\textbf{Single domain.} While the framework is domain-agnostic, our released benchmark covers movies only. Extending to other domains is straightforward given the typed-predicate scoring; cross-domain task suites and larger-$k$ runs are natural next steps.\\
\textbf{Statistical power.} We report 4 trials per task; 95\% bootstrap confidence intervals on \passhat{4} are non-trivial ($\pm$0.10--0.13). Distinguishing models with overlapping CIs would require either more trials per task (cost-bound) or more tasks per cell (annotation-bound).\\





\bibliographystyle{ACM-Reference-Format}
\bibliography{sample-base}

\end{document}